Valuation, Liquidity Price, and Stability of Cryptocurrencies

Carey Caginalp[1] and Gunduz Caginalp[2]

[1]Carnegie-Mellon University and University of Pittsburgh, carey_caginalp@alumni.brown.edu
[2]University of Pittsburgh, caginalp@pitt.edu, www.pitt.edu/~caginalp

The spectacular rise of Bitcoin's price has attracted the attention of many, including government regulators and speculators, in addition to those who wish to use a virtual currency, often with little trace or record [1]. On October 13, 2017, Bitcoin's market capitalization (number of Bitcoins multiplied by the trading price) surpassed both Goldman Sachs and Morgan Stanley as it catapulted past $96 billion, an increase of nine-fold over the previous year (see Figure 1). Governments have floundered as they scramble to control cryptocurrencies [2, 3, 4].

Many currencies and speculative instruments have evolved in modern times. However, we believe that the basic requirements for currencies and speculative assets are mutually exclusive. The former requires stability, namely, that tomorrow's purchasing power of the given currency should be nearly identical to today's. Prolonged stability, however, usually terminates the speculative interest in an asset. Thus far, Bitcoin, Ethereum, and some other cryptocurrencies seem to satisfy the conditions for speculation But, in our opinion, stability will not easily materialize.

We argue that an asset which has no value by traditional measures will tend to trade at a price that is determined largely by the fraction, $L$, of the amount of dollars available for the asset divided by the total number of units of the asset. This conclusion is deduced from mathematical modeling and economics experiments that we discuss below. Both strongly suggest that stability will be lacking, so the cryptocurrencies may simply be a mechanism for a transfer of wealth from the late-comers to the early entrants and nimble traders.

**Valuation and Bubbles.**

From an academic perspective, there is a growing sense that this is a new bubble, much like those before: the housing bubble of 2008, the Internet bubble of 1999, the South Sea bubble of 1720, and the Dutch tulip bulbs bubble of 1637 [6, 7]. As in the Internet bubble, the advent of a new technology is attractive to a large number of people who are blinded to the possible pitfalls of the investment.

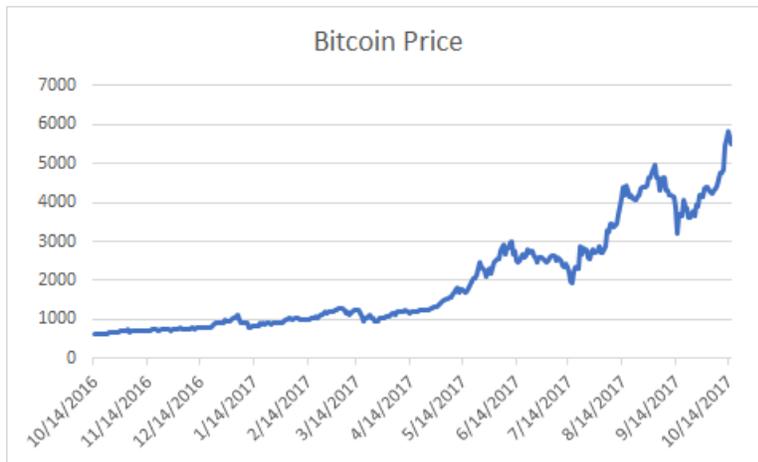

*Figure 1. The price of Bitcoin has risen approximately ninefold in the past year.*

As Bitcoin's price soars (see Figure 1), many turn to economics for an explanation. Cryptocurrencies are neither a proxy for a tangible asset such as an exchange traded fund investing in gold, nor a security, such as a common stock. There are well established methodologies for estimating the value of such instruments. For example, the science of measuring the value of a stock dates back to Ben Graham in the 1930's [8]. Traditional currencies have their own valuation mechanism based on economics and finance opportunities. Commodities such as gold and silver have a value based in part on industrial demand and utility.

The value of a cryptocurrency such as Bitcoin is unchartered territory in economics. It differs from the speculative manias of the past in that the sole purpose in owning the asset (unlike tulip bulbs, or internet stocks) is not speculation, but also as a vehicle to trade for tangible goods and services. Speculation is initially a secondary motivation, but could become dominant as prices soar. A question discussed below is why prices should move higher in the first place, given that some people choose to use Bitcoin as a currency.

**Insights from Mathematical Modeling and Experimental Economics**

Experimental asset markets, as well as asset flow differential equations that have been studied for the past three decades, offer insight into the valuation of a cryptocurrency. Vernon Smith (2002 Nobel Laureate), Gerry Suchanek, and Arlington Williams [9] introduced the basic "bubbles" experiments in which participants are endowed with cash and shares of a single asset that pays an expected value of 24 cents at the end of each of 15 trading periods to the holder at the end of that period. It becomes worthless after period 15. One can calculate the expected value of the asset as 360 cents initially and declining by 24 cents each period. This calculation can be regarded as the fundamental or intrinsic value of the asset. The trading price per share is established by the bid/ask matching process. In these experiments, which have been replicated many times, the trading price typically starts below the expected payout of 360 cents and begins to rise, culminating at a peak price far above the fundamental value.

Clearly, both trading price and fundamental value have units of dollars per share. There is, however, another quantity introduced by Caginalp and Balenovich [10] that also has these units. An examination of equilibrium led to the conclusion: "In the absence of clear information and attention to value, the price tends to gravitate to a natural value determined by the ratio of total cash to total quantity of asset" [10]. This is the liquidity price or value, that we denote by $L$.

This theoretical prediction was tested in several experiments [11, 12] in which all parameters and conditions were fixed while only the cash supply was altered. The first of these had an extremely simple structure: participants traded a single asset with an expected payout at the end, confirming the theoretical prediction. In experiments of the bubbles type format, it was found that each dollar of additional cash per share, raised the maximum price at the height of the bubble by about $1 per share (see Figure 2).

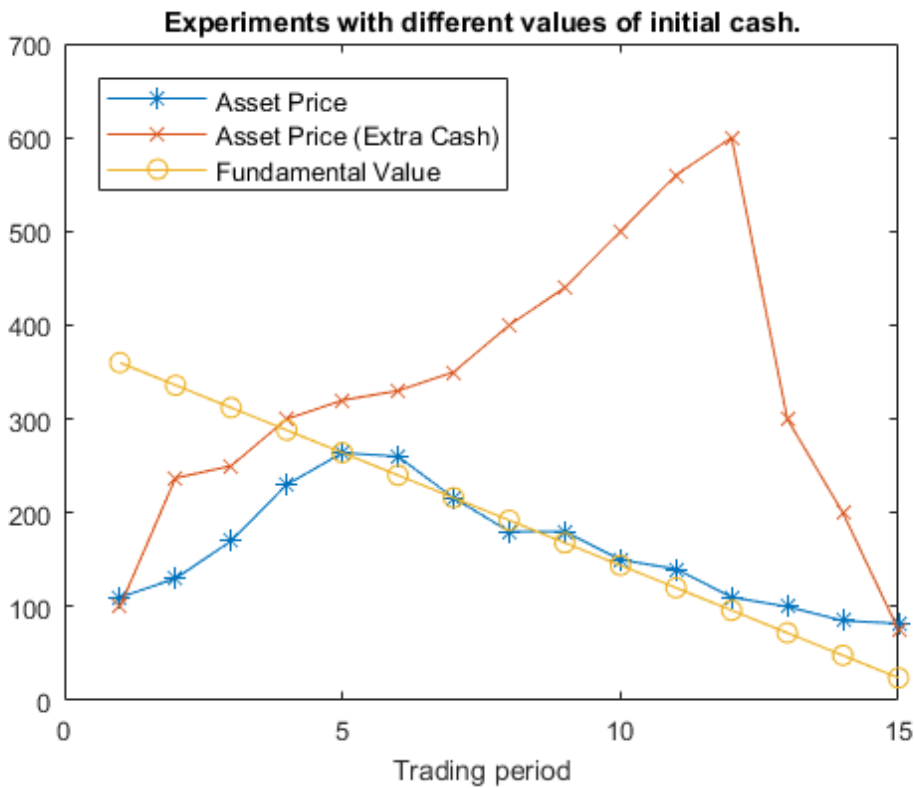

*Figure 2. The Caginalp, Porter and Smith experiments [12] use the classical "bubbles" design featuring one asset that pays a dividend throughout 15 trading periods. When the initial amount of cash per share, or liquidity value, is equal to the fundamental value, as in the first experiment above, there is no bubble and only slight fluctuations around the trading price due to randomness. When the initial cash is doubled, a large bubble is produced, with peak price close to the liquidity value.*

From the perspective of classical economics this result is surprising. Why should someone pay, say, $6.00 for an asset that will ultimately pay out $3.60, unless he can sell it to someone else at a higher price? But because the potential buyer would have access to the same information, game theory would suggest that no one would purchase it higher than the expected payout. One way to understand this phenomenon is that, in the absence of infinite arbitrage, the price is determined

"by the margin," not by the average of potential buyers. In other words, if there is a small supply of an asset, and many buyers, it is only the opinion of those at the fringe of the bell-shaped curve (who may be mistaken) that will determine the price. The middle (and perhaps wiser) part of the distribution has no role in it.

Another mechanism that is stipulated in the theory and was borne out in experiment and large-scale empirical work [14] is momentum, which contributes to the bubble in terms of an increase in the price that people are willing to pay for an asset, plus it draws in more cash from speculators. Ultimately though, these eager buyers turn into relentless sellers when prices start to fall.

**Application to Cryptocurrencies.**

In applying these theoretical and experimental ideas to Bitcoin, we assume for simplicity a single cryptocurrency and a group of investors who wish to use it for transactions that they cannot make using their national currency. How will the price evolve? The total demand for the cryptocurrency is the net dollar amount that these traders would like to use. The supply is the total number of units of the cryptocurrency. In the case of Bitcoin, the supply gradually increases with electronic "mining," meaning Bitcoins are generated for those solving a particular problem via computing. At first there will be some skepticism among the group needing the cryptocurrency, so the total number of dollars submitted for trading into the cryptocurrency would be small. As with Internet purchases in the 1990's, gradually people feel a greater sense of security, and there is increasing probability that someone who wishes to avoid the regulated financial system will use the cryptocurrency. After some time one can expect nearly all of those potential participants to be willing traders.

In other words, if $M$ is the dollar amount owned by the group seeking to bypass the traditional financial system, $p_c$ is the probability that the owner of each dollar is willing to use a cryptocurrency, and $N$ the number of Bitcoins, then the liquidity price would be $L = p_c M/N$. Currently, for Bitcoin, $N$ is approximately 16 million units (and capped at 21 million). One expects that $p_c$ will be increasing unless there are events that undermine confidence, such as the Mt. Gox hacking in February 2014 [15], announcements of regulatory crackdown, etc. The value of M is presumably difficult to estimate, but is probably more slowly varying in time compared to $p_c$. As of October 13, 2017 the trading price of Bitcoin is nearly $P = \$6,000$, giving it a market capitalization of $PN = (6000)(16 * 10^6)$ or $96 billion. If we estimate $L$ by $P$, then we obtain an estimate $p_c M = LN = 6000N$, i.e., $p_c M$ is also $96 billion. The worldwide demand for private transactions, namely, $M$, is presumably higher than this sum, suggesting that $p_c$ is much smaller than 1, even when the entire set of cryptocurrencies is taken into account with market cap at $170 billion. This suggests that only a fraction of the potential users are currently participating.[1]

**Absence of Valuation, Absence of Stability.**

The main forces behind the rising price of cryptocurrencies appear to be similar to many historical bubbles: increasing liquidity (as defined above) and momentum. But liquidity and momentum work in both directions. When prices begin to move down, the speculators who bought because of rising prices turn into determined sellers. If some holders of a cryptocurrency feel that their investment is not as secure as they thought, the liquidity price will diminish. This situation is not very different from the stock market bubbles we have seen, but there, the value-based investors stepped in at some stage to purchase bargains.

---

[1] Since the original submission on October 17, 2017, the price of Bitcoin (and with it the market cap) has more than tripled with recent trading near $20,000. Already in the intervening months, the analysis has been borne out.

With cryptocurrencies, the essential point is that the fundamental value is non-existent, and it is unlikely that the value-based buyers would step in at any price. After all, a stockholder of a healthy corporation is part owner of an entity that has tangible assets and ongoing business, and usually pays a dividend. What does ownership of a cryptocurrency assure for the holder?

The numerator in the liquidity price ($L$) has been defined as the total amount of dollars that are devoted to a cryptocurrency. This is likely to be significantly influenced by many factors including: (i) government actions, such as conventional currency restrictions that lead to greater demand for cryptocurrencies; (ii) government restrictions on trading cryptocurrencies that diminish the capital; (iii) losses due to hacking would diminish the appetite for the cryptocurrencies.

The denominator in the liquidity price also presents considerable uncertainty. Although Bitcoin is to be capped at 21 million units, it has had two offshoots, Bitcash and Bitgold, whereby closely related instruments are launched, effectively adding to the supply. Also, there is competition from other cryptocurrencies, for example, those offering more privacy.

The issue of the number of units and how they are issued illustrates the instabilities inherent in an instrument that is governed by a group that is not accountable to the owners of the currency. To which court does one appeal if there are actions that are detrimental to the owners' interests? And what is the contract?

As the experiments, modeling, and empirical studies show, momentum is reined in by traders focused on valuation. Momentum is thus likely to be a further destabilizing force in cryptocurrencies since there is an absence of value traders. Events that trigger a change in demand are likely to be followed by momentum buying or selling.

As the market capitalization of the cryptocurrencies increases to magnitudes that are significant in terms of the economy and financial institutions, further questions will arise. One such concern is leverage, i.e., buying on borrowed money. When leverage is involved, a sharp decline in the price of an asset can result in the insolvency of lending institutions, which in turn cannot pay their debt to other institutions. This cascading effect, that was central to the housing related crisis of 2008 [14] and the stock market crash of 1929 [5], has the potential to impact those who do not own the particular asset (see [1]). In their current form, the cryptocurrencies are unlikely to be stable enough to use as a currency. In the absence of a link to valuation, the price will likely be subject to the fickle winds of liquidity and momentum.

A mechanism that could make the connection with valuation would involve a structure based on the world's gross domestic product (GDP). The cryptocurrency would entail a contract (enforceable as with other financial instruments) that entitles the holder the right to redeem it as a specified fraction of the world's GDP in terms of a basket of major currencies or commodities.

[17] "Coindesk," October 2017. [Online]. Available: https://www.coindesk.com/price/.